\documentclass[10pt, conference]{IEEEtran}                               
\IEEEoverridecommandlockouts

\makeatletter
\def\ps@IEEEtitlepagestyle{%
  \def\@oddfoot{\mycopyrightnotice}%
  \def\@evenfoot{}%
}
\def\mycopyrightnotice{%
  \begin{minipage}{\textwidth}
  \justifying
  \justifying \scriptsize
  \noindent Copyright~\copyright~20xx IEEE. Personal use of this material is permitted. Permission from IEEE must be obtained for all other uses, in any current or future media, including reprinting/republishing this material for advertising or promotional purposes, creating new collective works, for resale or redistribution to servers or lists, or reuse of any copyrighted component of this work in other works by sending a request to pubs-permissions@ieee.org.
  \end{minipage}
}
\makeatother

\usepackage[utf8]{inputenc}
\usepackage[T1]{fontenc}
\usepackage{microtype}
\usepackage{graphicx}            

\usepackage{array}
\newcommand{\PreserveBackslash}[1]{\let\temp=\\#1\let\\=\temp}
\newcolumntype{C}[1]{>{\PreserveBackslash\centering}p{#1}}
\newcolumntype{R}[1]{>{\PreserveBackslash\raggedleft}p{#1}}
\newcolumntype{L}[1]{>{\PreserveBackslash\raggedright}p{#1}}
\usepackage{supertabular,booktabs}
\usepackage{graphicx}
\usepackage{multirow}

\makeatletter
\def\thickhline{%
  \noalign{\ifnum0=`}\fi\hrule \@height \thickarrayrulewidth \futurelet
   \reserved@a\@xthickhline}
\def\@xthickhline{\ifx\reserved@a\thickhline
               \vskip\doublerulesep
               \vskip-\thickarrayrulewidth
             \fi
      \ifnum0=`{\fi}}
\makeatother

\newlength{\thickarrayrulewidth}
\usepackage[mode=buildnew]{standalone} %
\usepackage{ragged2e}
\usepackage{float}
\graphicspath{{./fig/}}
\usepackage{parskip}
\usepackage{rotating} %
\usepackage{makecell}
\usepackage{enumerate}%
\usepackage{lettrine}
\usepackage{textcomp}
\usepackage[dvipsnames]{xcolor}
\RequirePackage[bookmarks, bookmarksopen=true, plainpages=false, pdfpagelabels]{hyperref}
\hypersetup{
colorlinks=true,
linkcolor=Blue,
filecolor=black,      
urlcolor=black,
citecolor = red,
}

\let\oldciteauthor=\citeauthor
\def\citeauthor#1{\hypersetup{citecolor=black}\oldciteauthor{#1}}
\let\oldcite=\cite
\def\cite#1{\hypersetup{citecolor=black}\oldcite{#1}}

\def\BibTeX{{\rm B\kern-.05em{\sc i\kern-.025em b}\kern-.08em
    T\kern-.1667em\lower.7ex\hbox{E}\kern-.125emX}}

\makeatletter
\let\old@makecaption=\@makecaption
\usepackage{subcaption}
\let\@makecaption=\old@makecaption
\makeatother

\usepackage{amsmath}
\usepackage{amssymb,amsfonts, mathtools}
\usepackage{bbm}
\usepackage{bm}
\usepackage{cuted}
\usepackage{etoolbox}
\usepackage{upgreek}
\AtBeginEnvironment{gather}{\setcounter{equation}{0}}
\newtheorem{lemma}{Lemma}
\newtheorem{remark}{Remark}
\newtheorem{definition}{Definition}

\newtheorem{corollary}{Corollary}
\newtheorem{assumption}{Assumption}
\allowdisplaybreaks %

\usepackage{algorithm,algpseudocode}
\usepackage{xpatch}
\makeatletter
\xpatchcmd{\algorithmic}
  {\ALG@tlm\z@}{\leftmargin\z@\ALG@tlm\z@}
  {}{}
\makeatother
\makeatletter
\newcommand*{\algrule}[1][\algorithmicindent]{%
  \hspace*{.2em}%
  \vrule %
  \hspace*{\dimexpr#1-.2em-.4pt}%
}

\newcommand{\StatePar}[1]{%
  \State\parbox[t]{\dimexpr\linewidth-\ALG@thistlm}{\strut #1\strut}%
}
\renewcommand{\ALG@beginalgorithmic}{\offinterlineskip}%

\newcount\ALG@printindent@tempcnta
\def\ALG@printindent{%
  \ifnum \theALG@nested > 0%
    \ifx\ALG@text\ALG@x@notext%
    \else
      \unskip
      \ALG@printindent@tempcnta=1
      \loop
        \algrule[\csname ALG@ind@\the\ALG@printindent@tempcnta\endcsname]%
        \advance \ALG@printindent@tempcnta 1
        \ifnum \ALG@printindent@tempcnta<\numexpr\theALG@nested+1\relax
      \repeat
        \fi
    \fi
}
\patchcmd{\ALG@doentity}{\noindent\hskip\ALG@tlm}{\ALG@printindent}{}{\errmessage{failed to patch}}
\makeatother
\algrenewcommand\algorithmicend{\strut\textbf{end}}
\algrenewcommand\algorithmicdo{\strut\textbf{do}}
\algrenewcommand\algorithmicwhile{\strut\textbf{while}}
\algrenewcommand\algorithmicfor{\strut\textbf{for}}
\algrenewcommand\algorithmicforall{\strut\textbf{for all}}
\algrenewcommand\algorithmicloop{\strut\textbf{loop}}
\algrenewcommand\algorithmicrepeat{\strut\textbf{repeat}}
\algrenewcommand\algorithmicuntil{\strut\textbf{until}}
\algrenewcommand\algorithmicprocedure{\strut\textbf{procedure}}
\algrenewcommand\algorithmicfunction{\strut\textbf{function}}
\algrenewcommand\algorithmicif{\strut\textbf{if}}
\algrenewcommand\algorithmicthen{\strut\textbf{then}}
\algrenewcommand\algorithmicelse{\strut\textbf{else}}

\algrenewcommand\algorithmicrequire{\strut\textbf{Input:}}
\algrenewcommand\algorithmicensure{\strut\textbf{Output:}}

\let\oldState\State
\renewcommand{\State}{\oldState\strut}

\newtheorem{theorem}{Theorem}

\newcommand{\ith}{i^{\text{th}}}

\usepackage{epsfig}

\usepackage{soul}

\usepackage{capt-of}
\definecolor{mplblue}{RGB}{31, 119, 180}

\usepackage{pgf, tikz}
\usetikzlibrary{shapes,arrows,external,calc, automata, arrows.meta,patterns} %
\usepackage{pgfplots}
\pgfplotsset{width=10cm,compat=1.9}
\tikzstyle{block} = [draw, fill=green!20, text centered, rectangle,
minimum height=0.8cm,minimum width=6em, align=right]

\begin{document}

\title{\fontsize{14}{14}\selectfont{\textbf{Consensus-Based Leader-Follower Formation Tracking for Control-Affine Nonlinear Multiagent Systems}\textsuperscript{*}}\thanks{\textsuperscript{*}This work received support in part from the Office of Naval Research (Grant No. N000141712622) and from a joint award from Microsoft Corporation, 1 Microsoft Way, Redmond, WA 98052, USA and the Maryland Robotics Center, 3156 Brendan Iribe Center, College Park, MD 20742, USA.}
\thanks{\textsuperscript{1}C. Enwerem and J.S. Baras are with the Department of Electrical \& Computer Engineering and the Institute for Systems Research, University of Maryland, College Park, MD 20742, USA. The work of C. Enwerem was supported by the Microsoft Diversity in Robotics and Autonomy PhD Fellowship.
Emails: \{\texttt{enwerem, baras}\}@umd.edu.}
}

\author{{Clinton Enwerem\textsuperscript{1}}
\and
{John S. Baras\textsuperscript{1}}}

\maketitle

\begin{abstract}                        
In the typical multiagent formation tracking problem centered on consensus, the prevailing assumption in the literature is that the agents' nonlinear models can be approximated by integrator systems, by their feedback-linearized equivalents, or by dynamics composed of deterministic linear and nonlinear terms. The resulting approaches associated with such assumptions, however, are hardly applicable to general nonlinear systems. To this end, we present consensus-based control laws for multiagent formation tracking in finite-dimensional state space, with the agents represented by a more general class of dynamics: control-affine nonlinear systems. The agents also exchange information via a leader-follower communication topology modeled as an undirected and connected graph with a single leader node. By leveraging standard tools from algebraic graph theory and Lyapunov analysis, we first derive a locally asymptotically stabilizing formation tracking law. Next, to demonstrate the effectiveness of our approach, we present results from numerical simulations of an example in robotics. These results --- together with a comparison of the formation errors obtained with our approach and those realized via an optimization-based method --- further validate our theoretical propositions.
\end{abstract}

\begin{IEEEkeywords}                        
multiagent systems, consensus, formation control,
control-affine nonlinear systems               
\end{IEEEkeywords}

\section{Introduction}

Multiagent formation tracking --- the task of controlling a group of dynamic units (or agents) to maintain a desired formation while following a reference trajectory --- has risen to become evidently one of the most popular topics in cooperative control and related fields, owing to its numerous practical applications. The formation \textit{tracking} problem is distinct from the nominal formation \textit{control} problem (where the focus instead is only on the agents converging to a desired formation) and has traditionally been tackled using ideas from fields such as learning theory, graph theory, optimization, and control theory, to name a handful. However, studies featuring methods from the aforementioned areas typically consider the agents' dynamic models as linear systems --- a common simplification from their general and more complex nonlinear representations. Within the consensus control domain particularly, several research articles \cite{dongConsensusBasedFormation2013,wangDistributedAdaptiveTimevarying2016} have studied the formation tracking problem, following the seminal work of \cite{olfati-saberConsensusCooperationNetworked2007}. These studies have been mostly adhoc, however, extending the aforementioned ideas to only time-invariant linear systems or systems with dynamics comprising deterministic linear and nonlinear terms. Unfortunately, the control schemes designed under these approximations are largely not generalizable and cannot be extended to the nonlinear case as a result. This significant research gap thus motivates the need for results applicable to a more general class of systems.

\subsection{Prior Work}

Results demonstrating the application of consensus to formation control for nonlinear systems began to appear fairly recently, starting in \cite{liuFormationControlMultiple2019}, which featured a consensus-based event-triggered control scheme, with the agents represented by nonlinear systems with a scalar control coefficient. Following that, in a closely-related work \cite{wangFormationTrackingConsensus2020}, the authors laid down consensus-based formation control laws for a class of nonlinear multiagent systems --- with models similar to those studied in \cite{liuFormationControlMultiple2019}. More recently, consensus-based formation control laws were given in \cite{https://doi.org/10.1002/rnc.6451}, but for nonlinear systems with linear-in-state drift terms. In our work, we derive original locally asymptotically stable consensus-based formation tracking laws for a more general class of nonlinear systems --- affine-in-control systems with a state-dependent drift term. Following \cite{huConsensusMultiAgentSystems2016, huConsensusClassDiscretetime2017}, we present general consensus control rules that guarantee asymptotic decay of the formation error of the multiagent system (MAS). In contrast to the aforementioned studies, we apply the agreement protocol to the problem of formation tracking, where the agents are modeled by general control-affine nonlinear systems.

\subsection{Main Contributions}
 We contribute the following to the state of the art in consensus-based formation tracking: (i) an original consensus-based formation tracking control scheme --- for control-affine nonlinear systems --- that guarantees asymptotic convergence of each agent's consensus error to the corresponding relative state, thus preserving the formation (Section \ref{sec:mainres}), and (ii) numerical formation tracking simulations that juxtapose the performance of our approach with that obtained by an optimization-based method (Section \ref{sec:simex}). 

\section{Notation \& Mathematical Preliminaries}
\label{sec:probform}

Throughout, we denote vectors and matrices with boldface font (e.g., $\bm{x}$ and $\bm{A}$), with corresponding elements in regular italicized font, i.e., $x_i$ denotes the $\ith$ element of the vector $\bm{x}$, while $A_{ij}$ is the element occupying the $\ith$ row and $j^{\text{th}}$ column of $\bm{A}$. $\bm{A}^T$ denotes the transpose of matrix $\bm{A}$, $\times$ (in the context of three-dimensional vectors) represents the cross product, and $\otimes$ is the standard Kronecker product operator. Unless otherwise noted, $\bm{I}_n$ is the $n\times n$ identity matrix, $\bm{\kappa}_n$ is the vector $\begin{bsmallmatrix} \kappa & \kappa & \dots & \kappa\end{bsmallmatrix}^T \in \mathbb{R}^n$, with $\kappa \in \mathbb{R}$, and $\lvert\lvert \bm{x} \rvert\rvert = (\sum_{i=1}^{n}(x_i)^2)^{\frac{1}{2}}$ is the vector Euclidean norm. $\text{\text{diag}}([\star_i]_{i\in{I}})$ is the $|{I}|\times|{I}|$ diagonal matrix, with ${I}$ an index set. The $\ith$ block, $\star_i$, can either be a scalar, vector, or matrix, and will be clear from the context. $\mathbb{R}_+$ is the set $\{\alpha \in \mathbb{R}\ \lvert \ \alpha>0\}$. To set the stage for the theoretical framework and problem formulation to follow, we now introduce several key lemmas, definitions, and theorems, to be invoked in the proofs to come.

 \begin{definition}[Control-Affine Nonlinear System]
     A control-affine nonlinear system is the system (here subscripted by $i$ to represent the $\ith$ agent for notational cohesion):
     \begin{subequations}
     \label{eq:contaffss}
         \begin{align}
            \label{eq:contaff}
            \dot{\bm{x}}_i &= f(\bm{x}_i) + g(\bm{x}_i)\bm{u}_i\\
            \label{eq:nlmodout}
            \bm{y}_i &= h(\bm{x}_i),
         \end{align}
     \end{subequations}
     where $\bm{x}_i \in \mathcal{X} \subset \mathbb{R}^n$, $\bm{u}_i \in \mathcal{U} \subseteq \mathbb{R}^m$, and $\bm{y}_i \in \mathbb{R}^q$ are, respectively, the state, control, and output vectors, with $m$ not necessarily equal to $q$. $f(\cdot)$, $g(\cdot)$, and $h(\cdot)$ are, respectively, the drift and control vector fields, and the output map --- with appropriate dimensions --- which are assumed to be sufficiently-smooth functions ($\mathcal{C}^n; n \geq 1$), with $f(\bm{0}_n) = \bm{0}_n$, so that $\bm{x}_i = \bm{0}_n$ is an open-loop equilibrium point of (\ref{eq:contaffss}), and $h(\bm{0}_n) = \bm{0}_q$. $\mathcal{X}$ is a compact set containing $\bm{0}_n$, while admissible control signals take values in the set, $\mathcal{U}$, of piecewise continuous and absolutely-integrable functions, i.e., the set $\{\bm{u}_i(t) \ | \ \int_{0}^{t}{|\bm{u}_i(\tau)|d\tau} < \infty \}$.
 \end{definition}

\begin{definition}[Graph Theory]
    An \textit{undirected}, finite, graph (hereafter graph) $\mathcal{G}$ is the tuple $(\mathcal{V}, \mathcal{E})$, with $\mathcal{V}$ equal to the non-empty set $\{ v_1, v_2, \dots, v_k\}$ of $k$ distinct elements, called \textit{nodes} or \textit{vertices} and $\mathcal{E}$ (the \textit{edge} set) $\subseteq$ $\mathcal{V}\times \mathcal{V} = \{\{v_i, v_j\} \ \lvert \ v_i, v_j \in \mathcal{V}\}$. The graph is \textit{weighted} if there exists a map $w: \mathcal{E} \mapsto \mathbb{R}$ that associates to each edge, a real number, denoted as $w_{ij}$. It is \textit{unweighted} otherwise. A \textit{path} is a sequence of distinct vertices $v_1, v_2, \dots, v_m$, where every consecutive pair of vertices (i.e., $v_i$ and $v_{i+1}$; $i = 1, \dots, m$) is joined by an edge in $\mathcal{E}$. A graph is \textit{connected} if every node in the graph is connected to every other node by a path; the graph is \textit{disconnected} otherwise. The neighborhood set, $N_i$, is the set $\{v_j \in \mathcal{V}\ \lvert \ \{v_i, v_j\} \in \mathcal{E}, \ j\neq i\}$. The graph adjacency and Laplacian matrices (denoted, respectively, as $\bm{\mathcal{A}}(\mathcal{G})$ and $\bm{\mathcal{L}}(\mathcal{G})$) associated with $\mathcal{G}$, are the matrices:
    \begin{subequations}
    \begin{align}
        \label{eq:adjmatr}
        \bm{\bm{\mathcal{A}}}(\mathcal{G}) &=[\mathcal{A}_{ij}], \ \text{where}\ 
        \mathcal{A}_{ij} = 
        \begin{dcases}
            1, \ \text{if} \ \{v_i, v_j\} \in \mathcal{E}\\
            0, \ \text{otherwise}
        \end{dcases}\\
        \label{eq:laplmatr}
        \bm{\mathcal{L}}(\mathcal{G}) &= [\mathcal{L}_{ij}], \ \text{where}\ 
        \mathcal{L}_{ij} = 
        \begin{dcases}
        \sum_{j \in N_i}^{}  \mathcal{A}_{ij}, \ \text{if} \ i=j\\
         -\mathcal{A}_{ij}, \ \text{otherwise}.
        \end{dcases}
    \end{align}
    \end{subequations}
\end{definition}

Hereafter, for brevity, we shall drop the $\mathcal{G}$ argument in the notations for the adjacency matrix and graph Laplacian.

\begin{figure}[htb]
\centering
\scalebox{0.8}{
 \begin{tikzpicture}[
            > = Stealth, %
            shorten > = 0pt, %
            auto,
            node distance = 2.5cm, %
            thick, %
            ]

        \tikzstyle{every state}=[
            draw = black,
            very thick,
            fill=gray!30,
            minimum size,
            inner sep=0pt
        ]

        \node[state] (l) {$a_L$};
        \node[state] (f1) [right of=l] {$a_{f_1}$};
        \node[state] (f2) [right of=f1] {$a_{f_2}$};
        \node[state] (fN) [right of =f2] {$a_{f_N}$};

        \path[] (f1) edge node {} (l);
        \path[] (f1) edge node {} (f2);
        \path[line width=1.25pt, dotted] (f2) edge node {} (fN);
    \end{tikzpicture}
    }
\caption{An unweighted line graph on $N+1$ vertices illustrating the leader-follower network structure under consideration. For clarity, the nodes have been renamed with the agent designations. Follower 1 ($a_{f_1}$), while alike to the other followers, is the only agent with the leader in its neighborhood set. Clearly, the network structure is a tree and not a complete graph.}
\label{fig:graphconc}
\end{figure}
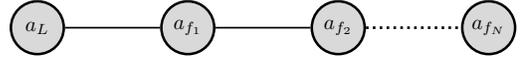

\begin{definition}[Leader-Follower Formation Tracking]
    Consider an interconnected system of $N+1$ identical agents --- each with dynamics described by (\ref{eq:contaff}) --- on the (connected) line graph depicted in Figure \ref{fig:graphconc}. For convenience, let the MAS comprise a single leader (with index $L$) and $N$ followers. The formation tracking problem is to synthesize controls, $\bm{u}_i$, under which the follower agents converge to states that respect the inter-agent distance constraints imposed by a specified formation rule, as the leader agent independently tracks a known reference trajectory, $\bm{x}_L^r$. Formally, taking $\bm{\xi}_i \in \mathbb{R}^n$ to be the desired (goal) state of the $\ith$ follower agent ($i = 1, 2, \dots, N$), given by the formation specification, and $\bm{x}_i$ to be the $\ith$ agent's actual state, the formation tracking problem is to find a $\bm{u}_i$ that drives the $\ith$ follower agent so that
    \begin{equation}
        \label{eq:constreq}
        ||\bm{x}_j(t) - \bm{x}_i(t)|| = \lvert\lvert\bm{d}_{ij}\rvert\rvert = \delta_{ij}, \ \forall \ j \ \in N_i,\ \forall \ t\ge 0,
    \end{equation}
    and a $\bm{u}_L$ that drives the leader such that
    \begin{equation}
        \label{eq:leaderr0}
        \lim_{t \rightarrow \infty}{\lvert\lvert \bm{x}_L(t) - \bm{x}_L^r(t)\rvert\rvert = 0},
    \end{equation}
\end{definition}
where $\bm{d}_{ij} = (\bm{\xi}_j - \bm{\xi}_i) \in \mathbb{R}^n$ is a constant vector, with norm ${\delta}_{ij} \in \mathbb{R}_+$. 

Recent studies \cite{barasFreshLookNetwork2014} have argued for the representation of the interaction in networked MAS using a three-layer multi-graph model as opposed to the ubiquitous single-layer model that often portrays only communication or information exchange. For completeness, therefore, we note here that Figure \ref{fig:graphconc} depicts the connection of the agents on the communication layer; the collaboration and information-sharing layers are taken to be subsumed in the network. 

\begin{theorem}[Mesbahi and Egerstedt \cite{mesbahiGraphTheoreticMethods2010}]
\label{theo:mesbahi1}
    For a connected goal formation, encoded by the graph, $\mathcal{G}_f = (\mathcal{V}, \mathcal{E}_f)$, and a goal location set $\mathcal{X}_f = \{\bm{\xi}_1, \bm{\xi}_2, \dots, \bm{\xi}_N\}$, where $\bm{\xi}_i$ is the goal location of the $\ith$ agent, the following formation control scheme
    \begin{equation}
        \label{eq:meshbahiformcont}
        \dot{\bm{x}}_i(t) = -\sum_{j \in N_i}^{}(\bm{x}_i (t) - \bm{x}_j (t)) - (\bm{\xi}_i - \bm{\xi}_j)
    \end{equation}
    will drive the MAS so that the agents converge to a constant displacement of the target positions, $\bm{\tau}$, i.e., for all agents,
        $\bm{x}_i(t) - \bm{\xi}_i \rightarrow \bm{\tau}  \text{ as } t \rightarrow \infty$.
\end{theorem}

\begin{assumption}
\label{asm:contauto}
    The system in (\ref{eq:contaffss}) is autonomous, hence associated notions of Lyapunov stability apply.
\end{assumption}

\begin{assumption}
\label{asm:gconnect}
    The network structure of the MAS is encoded by an unweighted, connected, and static graph, i.e., neither $\mathcal{V}$ nor $\mathcal{E}$ are time-dependent.
\end{assumption}

\begin{assumption}
    \label{asm:distfreeperfstate}
    The agents are taken to be represented by models with exact state information, so we do not consider the influence of noise or exogenous disturbances.
\end{assumption}

\begin{assumption}
    \label{asm:leadcont}
    The leader is driven by an independent control ($\bm{u}_{\text{track}}$), so that it asymptotically tracks its reference, $\bm{x}_L^r(t) \ \forall \ t \in [0, T]$, with $T$ finite.
\end{assumption}

\begin{assumption}
    \label{asm:plmi}
    There exists a positive definite matrix $\bm{P} \in \mathbb{R}^{n\times n}$ such that the following inequality holds:
    \begin{align}
    \label{eq:pineq1}
        &(\bm{x}_i - \bm{x}_j)^T\bm{P}(f(\bm{x}_i) - f(\bm{x}_j)) \leq 0\nonumber\\
        &\ \forall \ i, j = 1, 2, \dots, N \ \text{and} \ t \ge 0,
    \end{align}
    which places bounds on the relative inter-agent states of the unforced nonlinear dynamics for the multiagent system.
\end{assumption}

\begin{assumption}[\cite{huConsensusMultiAgentSystems2016}, Remark 1]
    \label{asm:mmatrix}
    There exists a positive definite $N\times N$ matrix $\bm{M}$ such that:
    \begin{equation}
        \label{eq:mmatrixeq}
        \bm{M}\bm{\mathcal{L}} = \bm{I}_N,
    \end{equation}
    which translates to an existence of the graph Laplacian's inverse and hence, the connectivity of the associated graph (see Lemma \ref{lem:lrealsym}).
\end{assumption}

\section{Main Result: Consensus-based Formation Maintenance Algorithm}
\label{sec:mainres}
Suppose Assumption \ref{asm:leadcont} holds. We define the $\ith$ consensus error as:
\begin{equation}
    \label{eq:conserr}
    \bm{e}_i = \sum_{j \in N_i}^{}{\mathcal{A}_{ij}(\bm{x}_i -\bm{x}_j)}.
\end{equation}
Then, from Theorem \ref{theo:mesbahi1}, the inter-agent distance constraints (together with the tracking requirement) can be achieved by respectively selecting the following control law for the $\ith$ follower and leader:
\begin{align} 
    \label{eq:ithcontlaw}
    &\bm{u}_{i}(t) = -g^T(\bm{x}_i(t))\bm{P}(\bm{e}_i(t) + \bm{d}_{i}),\ i = 1,2, \dots, N,\\
    \label{eq:leadlaw}
    &\bm{u}_{L}(t) = -g^T(\bm{x}_L(t))\bm{P}(\bm{e}_L(t) + \bm{d}_{L}) + \bm{u}_{\text{track}},
\end{align}
where $\bm{d}_i = \sum_{j \in N_i}^{}\bm{d}_{ij}$, and $\bm{P}$ satisfies (\ref{eq:pineq1}).
\begin{remark}
    We will show that, with (\ref{eq:ithcontlaw}), the follower agents reach consensus (asymptotically) on the inter-agent distances specified by the formation rule. That is, we can think of the consensus error as converging to the prescribed agent-to-agent distances (as opposed to zero in the nominal consensus case), which implies asymptotic decay of the formation error. Also notice that, since $N_L = \{a_{f_1}\}$, the leader is driven by a control signal (\ref{eq:leadlaw}) that balances the tracking and formation maintenance requirements.
\end{remark}
To  prove the validity of this result, we first define the following ensemble notation (for the follower-only network):
\begin{equation}
    \label{eq:blockx}
    \bm{x} = 
    \begin{bmatrix}
      \bm{x}_1^T & \bm{x}_2^T & \dots & \bm{x}_{N}^T 
    \end{bmatrix}^T \in \mathbb{R}^{{N}n},
\end{equation}
\begin{equation}
    \label{eq:blocke}
    \bm{e} = 
    \begin{bmatrix}
      \bm{e}_1^T & \bm{e}_2^T & \dots & \bm{e}_{N}^T 
    \end{bmatrix}^T \in \mathbb{R}^{{N}n},
\end{equation}
\begin{equation}
    \label{eq:blocku}
    \bm{u} = 
    \begin{bmatrix}
      \bm{u}_1^T & \bm{u}_2^T & \dots & \bm{u}_{N}^T 
    \end{bmatrix}^T \in \mathbb{R}^{{N}m},
\end{equation}
\begin{equation}
    \label{eq:blockg}
    \bm{G}(\bm{x}) = 
    \text{\text{diag}}\Big(\begin{bmatrix}{g(\bm{x}_1)} & {g(\bm{x}_2)} & \dots & {g(\bm{x}_{N})} \end{bmatrix}\Big) \in \mathbb{R}^{{N}n \times {N}m},
\end{equation}
\begin{equation}
    \label{eq:blockf}
    \bm{F}(\bm{x}) = 
    \begin{bmatrix}
      f^T(\bm{x}_1) & f^T(\bm{x}_2) & \dots & f^T(\bm{x}_{N})
    \end{bmatrix}^T \in \mathbb{R}^{{N}n},
\end{equation}
and
\begin{equation}
    \label{eq:blockd}
    \bm{d} = 
    \begin{bmatrix}
      \bm{d}_{1}^T & \bm{d}_{2}^T & \dots & \bm{d}_{{N}}^T 
    \end{bmatrix}^T \in \mathbb{R}^{{N}n}.
\end{equation}
We can then write (\ref{eq:ithcontlaw}) as:
\begin{equation}
    \label{eq:ublockcons}
    \bm{u}(t) = -\bm{G}^T(\bm{x}(t))(\bm{I}_N\otimes \bm{P})(\bm{e}(t) + \bm{d}),
\end{equation}
with the equation
\begin{equation}
    \label{eq:colldyn}
    \dot{\bm{x}} = \bm{F}(\bm{x}) + \bm{G}(\bm{x})\bm{u}
\end{equation}
representing the ensemble dynamics of the MAS in block notation. With this block formulation, it is easy to see that $\bm{e}$ in (\ref{eq:blocke}) satisfies the following dynamical equation:
    \begin{equation}
    \label{eq:ealtdyneq}
    \dot{\bm{e}} = (\bm{\mathcal{L}}\otimes \bm{I}_n)\dot{\bm{x}} = (\bm{\mathcal{L}}\otimes \bm{I}_n)(\bm{F}(\bm{x}) + \bm{G}(\bm{x})\bm{u}).
\end{equation}
Next, we introduce key lemmas (mostly adapted from \cite{huConsensusMultiAgentSystems2016}) which will be useful for proving the results to follow.
\begin{lemma}[\cite{mesbahiGraphTheoreticMethods2010}]
    \label{lem:lrealsym}
    For a connected leader-follower graph on $N+1$ vertices with one leader and $N$ followers, if $\bm{\mathcal{L}} \in \mathbb{R}^{N\times N}$ is the Laplacian corresponding to the follower network, then $\bm{\mathcal{L}}$ is real, symmetric, and positive definite, with its eigenvalues related by: $0 < \lambda_1 \le \lambda_2 \le \dots \le \lambda_N$.
\end{lemma}
\begin{lemma}[Schur Complement Lemma \cite{boydLinearMatrixInequalities1994}]
    \label{lem:schur}
    For any real and symmetric matrix $\bm{K} = \begin{bsmallmatrix}
        \bm{K}_{11} & \bm{K}_{12}\\
        \bm{K}_{21} & \bm{K}_{22}
    \end{bsmallmatrix}$,
    with $\bm{K}_{21} = \bm{K}^T_{12}$, the following statements are equivalent: (i)  $\bm{K} < 0$; (ii) $\bm{K}_{11} < 0$, $\bm{K}^T_{22} - \bm{K}^T_{12}\bm{K}_{11}^{-1}\bm{K}_{12} < 0$; (iii) $\bm{K}_{22} < 0$, $\bm{K}^T_{11} - \bm{K}_{12}\bm{K}_{22}^{-1}\bm{K}^T_{12} < 0$.
\end{lemma}
\begin{lemma}[\cite{huConsensusMultiAgentSystems2016}]
\label{lem:dlapab}
    For a connected graph $\mathcal{G}$ with Laplacian $\bm{\mathcal{L}}$ and adjacency matrix $\bm{\mathcal{A}}$ = [$\mathcal{A}_{ij}$], for any $\bm{h} = [{\bm{h}}_1^T {\bm{h}}_2^{T} \dots {\bm{h}}_N^T]^T$ and $\bm{k} = [{\bm{k}}_1^T {\bm{k}}_2^{T} \dots {\bm{k}}_N^T]^T$ in $\mathbb{R}^{Nn}$,
    \begin{equation}
        2\bm{h}^T(\bm{\mathcal{L}}\otimes \bm{I}_n)\bm{k} = \sum_{i=1}^{N}\sum_{j=1}^{N}\mathcal{A}_{ij}(\bm{h}_i -\bm{h}_j)^T(\bm{k}_i - \bm{k}_j).
    \end{equation}
\end{lemma}
\begin{lemma}
    \label{lem:eliineq}
    If Assumptions \ref{asm:gconnect} and \ref{asm:plmi} hold, then:
    \begin{equation}
        \label{eq:eplusdineq}
        (\bm{e}+\bm{d})^T(\bm{I}_N \otimes \bm{P})\bm{F(\bm{x})} \le 0.
    \end{equation}
\end{lemma}
\begin{IEEEproof}(Motivated by \cite{huConsensusMultiAgentSystems2016})
    From (\ref{eq:blockf}), we can write 
    \begin{align}
        &(\bm{I}_N \otimes \bm{P})\bm{F(\bm{x})}\nonumber\\ 
        &= \begin{bmatrix}
      f^T(\bm{x}_1)\bm{P} & f^T(\bm{x}_2)\bm{P} & \dots & f^T(\bm{x}_N)\bm{P}
    \end{bmatrix}^T\\
    &\triangleq \bm{F}^{\bm{P}}(\bm{x}).
    \end{align}

    From the left-hand side of (\ref{eq:eplusdineq}), it follows that:
    \begin{align}
        \label{eq:eplusdasx}
        &(\bm{e}+\bm{d})^T(\bm{I}_N \otimes \bm{P})\bm{F(\bm{x})}\nonumber\\
        &= (\bm{e}+\bm{d})^T\bm{F}^{\bm{P}}(\bm{x})\nonumber\\
        &= (\bm{x}^T(\bm{\mathcal{L}}\otimes\bm{I}_n) + \bm{d}^T)\bm{F}^{\bm{P}}(\bm{x})
    \end{align}

    Invoking Lemma \ref{lem:dlapab}, we can write (\ref{eq:eplusdasx}) as:
    \begin{align}
    \label{eq:eplusdlst}
            &(\bm{e}+\bm{d})^T(\bm{I}_N \otimes \bm{P})\bm{F(\bm{x})}\nonumber\\ 
            &= \frac{1}{2}\sum_{i=1}^{N}\sum_{j=1}^{N}\mathcal{A}_{ij}(\bm{x}_i-\bm{x}_j)^T\bm{P}(f(\bm{x}_i)-f(\bm{x}_j))\nonumber\\
            & \quad +\sum_{j\in N_i}\bm{d}_{ij}^T\bm{P}(f(\bm{x}_i)-f(\bm{x}_j)),
    \end{align}

    which is $\le 0$ by Assumption \ref{asm:plmi} and since $\mathcal{A}_{ij} \ge 0 \ \forall \ i,j = {1,2, \dots, N}$ under Assumption \ref{asm:gconnect}.
\end{IEEEproof}

\begin{theorem}
Suppose Assumptions \ref{asm:gconnect} and \ref{asm:plmi} hold. With the control defined in (\ref{eq:ithcontlaw}), the followers in the MAS will eventually converge to states respecting (\ref{eq:constreq}).
\end{theorem}

\begin{IEEEproof}(Motivated by \cite{huConsensusMultiAgentSystems2016})
    We begin the proof by selecting the Lyapunov function candidate:
    \begin{equation}
        \label{eq:lyapfxn}
        V = \frac{1}{2}(\bm{e}(t) + \bm{d})^T(\bm{M}\otimes\bm{P}) (\bm{e}(t) + \bm{d}),
    \end{equation}
    where $\bm{P}$ and $\bm{M}$ are as previously defined in Assumptions (\ref{asm:plmi}) and (\ref{asm:mmatrix}), respectively. Then, omitting the time and $\bm{x}$ arguments for conciseness, we can write the time derivative of $V$ along the trajectories of the closed-loop system as:
    \begin{align}
    \dot{V} &= (\bm{e} + \bm{d})^T(\bm{M}\otimes\bm{P})\dot{\bm{e}}\\
            &= (\bm{e} + \bm{d})^T(\bm{M}\otimes\bm{P})(\bm{\mathcal{L}}\otimes \bm{I}_n)(\bm{F} + \bm{G}\bm{u})\\
            &= (\bm{e} + \bm{d})^T(\bm{M}\bm{\mathcal{L}}\otimes\bm{I}_n)(\bm{I}_N\otimes \bm{P})(\bm{F} + \bm{G}\bm{u}).
    \end{align}
    By (\ref{eq:ublockcons}) and under Assumption \ref{asm:mmatrix}, we can then write:
    \begin{align}
        \label{eq:lyapfxnderiv}
        \dot{V} &= (\bm{e} + \bm{d})^T(\bm{I}_N\otimes \bm{P})(\bm{F} + \bm{G}\bm{u})\\
        \dot{V} &= (\bm{e} + \bm{d})^T(\bm{I}_N\otimes \bm{P})\bm{F}\nonumber\\
                \label{eq:vdotexpand}
                & - (\bm{e} + \bm{d})^T(\bm{I}_N\otimes \bm{P})\bm{G}\bm{G}^T(\bm{I}_N\otimes \bm{P})(\bm{e} + \bm{d}).
    \end{align}
    By Lemma \ref{lem:eliineq} and since the second term of (\ref{eq:vdotexpand}) is always negative, it follows that $\dot{V} \le 0$. Thus, the Lyapunov function (\ref{eq:lyapfxn}) will decrease along the trajectories of the closed-loop system, which implies (local) asymptotic stability of the equilibrium point of (\ref{eq:ealtdyneq}). Since, with the change of variables $\bm{z}(t) = \bm{e}(t) + \bm{d}$, it is easy to see that $\dot{\bm{z}} \equiv \dot{\bm{e}}$. Thus, from (\ref{eq:vdotexpand}) and (\ref{eq:ealtdyneq}), it follows that $\bm{z}(t) \rightarrow \bm{0}_n$ as $t \rightarrow \infty \implies ||\bm{e}(t)|| \rightarrow ||\bm{d}||$ as $t \rightarrow \infty$, thus completing the proof.
\end{IEEEproof}
\begin{corollary}
Define the formation error for (\ref{eq:colldyn}) as $\bm{f}(t)$ = $\bm{d} - \bm{e}(t)$. Since $||\bm{e}(t)|| \rightarrow ||\bm{d}||$ as $t \rightarrow \infty$, it immediately follows that $||\bm{f}(t)|| \rightarrow ||\bm{d} - \bm{d}|| = 0$, as $t \rightarrow \infty$, thus confirming asymptotic decay of the MAS's formation error.
\end{corollary}
\begin{remark}
    For the popular case of the formation tracking problem, where the agents' models are linear time-invariant systems --- equivalent to setting $f(\bm{x}_i) = \bm{A}\bm{x}_i$ and $g(\bm{x}_i) = \bm{B}$ in (\ref{eq:contaff}), with $\bm{A}$ and $\bm{B}$ constant matrices of appropriate dimensions (see \cite{https://doi.org/10.1002/rnc.6451} for example) --- the prevailing approach (assuming full controllability of the linear system) is to define $\bm{u}_i$ in terms of some control gain matrix $\bm{K}$, solve for a positive definite matrix $\bm{P}$ that satisfies the linear system's algebraic Riccati equation, and express $\bm{K}$ in terms of $\bm{P}$, e.g., $\bm{K} = -\bm{B}^{T}\bm{P}^{-1}$. The interested reader can consult \cite{liCooperativeControlMultiAgent2017} for a detailed treatment on the topic.
\end{remark}
\begin{theorem}
    Suppose a $\bm{P}$ satisfying (\ref{eq:pineq1}) exists. Then, it must be the case that $\bm{P}$ also satisfies the following linear matrix inequality (LMI)
    \begin{equation}
    \label{eq:nlmi}
    \begin{bmatrix}
        \bm{0}\in \mathbb{R}^{Nn\times Nn} &\Bar{\bm{P}}_{{Nn}}^{\frac{1}{2}}\\
       \Bar{\bm{P}}_{{Nn}}^{\frac{1}{2}} & -\bm{\Delta}f\bm{D}^T
    \end{bmatrix} \le 0,
    \end{equation}
    where $\bm{D} \in \mathbb{R}^{Nn\times N}$ is the matrix
    \begin{equation}
    \label{eq:dblockmatrix}
        [d_{ij}], \ \text{\normalfont with} \ d_{ij} = 
        \begin{cases}
            \bm{0}_n, \ \text{\normalfont if} \ i = j\\
             \bm{d}_{ij}, \ \text{\normalfont otherwise,}
        \end{cases}
    \end{equation}
    \begin{equation}
    \label{eq:pnneq}
        \Bar{\bm{P}}_{{Nn}} = \bm{I}_N \otimes \bm{P}^{-1}, \ \text{\normalfont and}
    \end{equation}
    $\bm{\Delta}f \in \mathbb{R}^{Nn\times N}$ is the matrix
    \begin{equation}
        \label{eq:deltafblockmatrix}
        [\delta f_{ij}], \ \text{\normalfont with} \ \delta f_{ij} =
        \begin{cases}
        \bm{0}_n, \ \text{\normalfont if} \ i = j\\
        f(\bm{\xi}_i) - f(\bm{\xi}_j), \ \text{\normalfont otherwise.}
        \end{cases}
    \end{equation}
\end{theorem}
\begin{IEEEproof}
    We begin the proof by noting (from (\ref{eq:pineq1})) that $\bm{P}$ must satisfy:
    \begin{align}
        &(\bm{\xi}_i - \bm{\xi}_j)^T\bm{P}(f(\bm{\xi}_i) - f(\bm{\xi}_j)) \leq 0\\
        \label{eq:equivpineq1}
        \implies & \bm{d}_{ij}^T\bm{P}(f(\bm{\xi}_i) - f(\bm{\xi}_j)) \leq 0. 
    \end{align}
    By block diagonalization (as in (\ref{eq:dblockmatrix}) and (\ref{eq:deltafblockmatrix})) and using the Kronecker product, we can show that (\ref{eq:equivpineq1}) is equivalent to
    \begin{equation}
        \label{eq:blockpineq}
        \bm{D}^T(\bm{I}_N \otimes \bm{P})\bm{\Delta}f \le 0,
    \end{equation}
    with $\bm{D}$ and $\bm{\Delta}f$ already defined. Invoking Lemma \ref{lem:schur}, it is straightforward to show that (\ref{eq:blockpineq}) can be expressed as (\ref{eq:nlmi}), which ends the proof.   
\end{IEEEproof}
\begin{remark}
\label{rem:findp}
    In the preceding theorem, a change of variables (\ref{eq:pnneq}) was necessary, to simplify the notation and allow for an immediate invocation of the Schur complement lemma. However, it is trivial to show that the original matrix of interest ($\bm{P}$) can be readily obtained from the uppermost-left $n \times n$ block of $\Bar{\bm{P}}_{{Nn}}^{-1}$. We are also guaranteed a factorization of the form $\Bar{\bm{P}}_{{Nn}}= \Bar{\bm{P}}_{{Nn}}^{\frac{1}{2}}\Bar{\bm{P}}_{{Nn}}^{\frac{1}{2}}$, since $\Bar{\bm{P}}_{{Nn}} \succ 0$ and $\otimes$ preserves positive definiteness. 
\end{remark}

We now present the following algorithm for formation maintenance. $T_{\Sigma}$ and $\bm{x}_i(0)$ are the total number of unity-spaced time steps (in [$0, T$], for a given interval) and the initial state of the $\ith$ follower agent, respectively.
\begin{algorithm}[H]
        \caption{Formation Maintenance Algorithm}
        \label{alg:formmain}
        \begin{algorithmic}   
        \State \textbf{Inputs: } $ \bm{D}, \bm{\Delta}f, N, N_i, \bm{\mathcal{A}}, \delta t, g(\cdot), f(\cdot)$, $\bm{x}_i(0)$, $\bm{x}_L(k)$\;
        \State \textbf{Outputs }: $\bm{P}, \bm{e}_i(k), \bm{x}_i(k); \ i = {1, 2, \dots, N} $\;
        \State Solve (\ref{eq:nlmi}) for $\Bar{\bm{P}}_{{Nn}}$ and obtain $\bm{P}$ (see Remark \ref{rem:findp})\;
        \For{$k\gets0$ \text{\textbf{ to }} $T_\Sigma$}    
        \For{$i\gets1$  \text{\textbf{ to }} $N$}    
            \State Calculate $\bm{e}_i$ from (\ref{eq:conserr})\;
            \State Substitute $\bm{P}$ and $\bm{e}_i$ from above steps in (\ref{eq:ithcontlaw})\;
            \State $\bm{x}_i(k+1) \gets f(\bm{x}_i(k)) + g(\bm{x}_i(k))\bm{u}_i(k)$
        \EndFor
        \EndFor
        \end{algorithmic}
\end{algorithm}

\section{Formation Specification}
\label{sec:formspec}
This section follows our previous work \cite{enweremDistributedOptimalFormation2023a}; however, here, the position of $\ith$ agent is in $\mathbb{R}^3$ and not the plane. As in  \cite{enweremDistributedOptimalFormation2023a}, we consider a triangular formation (with one leader and three followers) and also assume this formation to be feasible and rigid (in the sense of \cite{mesbahiGraphTheoreticMethods2010}, \S6), and invariant to homogeneous transformations, with appropriate dimensions. 

\section{Simulation Example}
\label{sec:simex}
As an example, consider the following dynamics for a six degrees-of-freedom quadrotor, adapted from \cite{misinLPVMPCControl2020} and described here (for brevity) using the Newton-Euler equations:%
\begin{equation}
    \label{eq:quadmodel}
    \resizebox{0.45\columnwidth}{!}{%
     $\begin{drcases}
        \dot{\mathbf p}_{\mathcal{I}} &= \bm{v}_{\mathcal{I}}\\
        \dot{\bm{v}}_{\mathcal{I}} &= \frac{1}{m}\bm{f}_{\mathcal{I}}\\
        \dot{\bm{\zeta}}_{\mathcal{I}} &= \bm{T}_{_{\mathcal{B}}}^{_{\mathcal{I}}}\bm{\omega}_{\mathcal{B}}\\
        \dot{\bm{\omega}}_{\mathcal{B}} &= {\bm{J}}^{-1}(\bm{\tau}_\mathcal{B}-\bm{\omega_{\mathcal{B}}} \times \bm{J\omega_{\mathcal{B}}}),
        \end{drcases}$%
        }
\end{equation}
where the variables subscripted by ${\mathcal{I}}$ represent quantities in the inertial frame, while those with a ${\mathcal{B}}$ subscript pertain to the quadrotor's body frame. $\bm{p}_{\mathcal{I}} = \begin{bsmallmatrix} x & y & z \end{bsmallmatrix}^T$ and $\bm{v}_{\mathcal{I}} = \begin{bsmallmatrix} {v_x} & {v_y} & {v_z}\end{bsmallmatrix}^T$ are, respectively, vectors of the $X$, $Y$, and $Z$ positions and linear velocities of the quadrotor, $\bm{\zeta}_{\mathcal{I}} = \begin{bsmallmatrix} \phi & \theta & \psi \end{bsmallmatrix}^T$ is the quadrotor's attitude vector (of roll, pitch, and yaw angles), while $\bm{\omega}_{\mathcal{B}} = \begin{bsmallmatrix} p & q & r \end{bsmallmatrix}^T$ is the vector of corresponding body-frame attitude rates. The parameters $m$, $\bm{J} = \text{\text{diag}}(\begin{bsmallmatrix}{I_{xx}} & {I_{yy}} & {I_{zz}} \end{bsmallmatrix})$, and $g$, represent the mass, inertia matrix, and acceleration due to gravity, respectively. Finally, $\bm{f}_{\mathcal{I}} = \begin{bsmallmatrix} {f_{\mathcal{I}}^{t}} & {f_{\mathcal{I}}^{t}} & {f_{\mathcal{I}}^{t}} \end{bsmallmatrix}^T$ and $\bm{\tau}_{\mathcal{B}} = \begin{bsmallmatrix} {\tau_x} & {\tau_y} & {\tau_z} \end{bsmallmatrix}^T$ are, respectively, vectors of the total thrust force ($f^t_{\mathcal{I}}$) and torques (about the roll, pitch, and yaw axes) applied to the quadrotor, while $\bm{T}_{_{\mathcal{B}}}^{_{\mathcal{I}}} \in \mathbb{R}^{3\times 3}$ is a matrix that transforms the body-frame angular velocities to the inertial frame.
\begin{figure*}[t]
     \centering
     \begin{subfigure}[b]{0.48\textwidth}
        \centering
    \includegraphics[trim=90pt 30 60 45,clip, width=0.8\columnwidth]{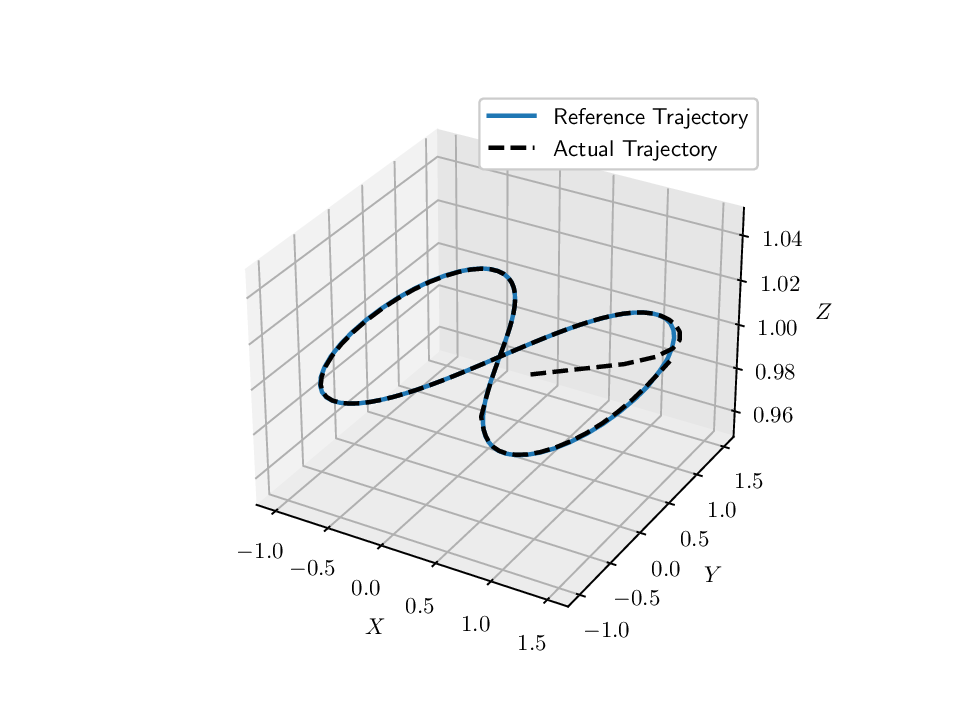}
        \caption{\label{fig:leadertraj}}
     \end{subfigure}
     \hfill
     \begin{subfigure}[b]{0.48\textwidth}
        \centering
        \includegraphics[trim=90pt 30 60 45,clip, width=0.8\columnwidth]{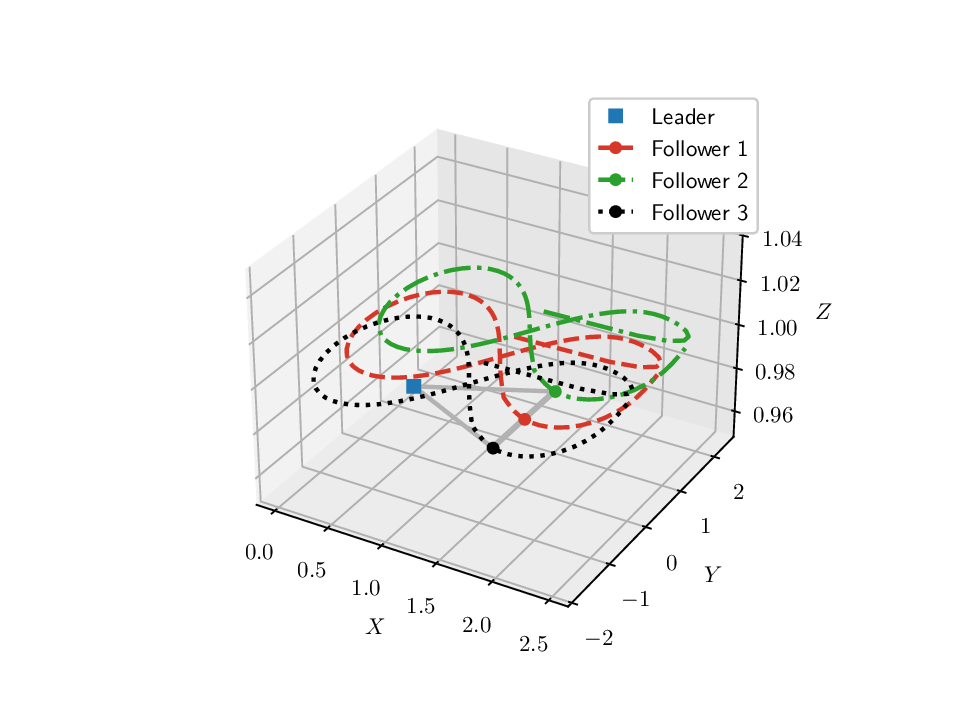}
        \caption{\label{fig:consall}}
     \end{subfigure}
     \caption{(\subref{fig:leadertraj}) Leader loop trajectory vs. reference. (\subref{fig:consall}) Leader trajectory tracking with triangular formation maintenance.}
\end{figure*}

It has been shown in \cite{mutohControlQuadrotorUnmanned2016} that the model in (\ref{eq:quadmodel}) can be expressed as a control-affine nonlinear system (\ref{eq:contaff}), but we will skip the details to adhere to page limits. 
To simulate the quadrotor dynamics, we select the parameters of the Crazyflie 2.1 quadcopter (Table \ref{tab:quadparam}), adapted from \cite{forsterSystemIdentificationCrazyflie2015a}.
\begin{table}[t]
    \caption{Quadrotor Parameters}
    \label{tab:quadparam}
    \centering
    \resizebox{\columnwidth}{!}{%
    \begin{tabular*}{0.6\textwidth}{@{\extracolsep{\fill}}*{6}{c}}
    \thickhline
    \multirow{2}{*}{\textbf{Parameter}} & \textbf{Mass} & \textbf{Arm Length} & \bm{ $I_{xx}$} & \bm{$I_{yy}$ }& \bm{$I_{zz}$}\\
    \cline{2-2}\cline{3-3}\cline{4-6}
    & kg & m & \multicolumn{3}{c}{$10^{-6}$ kg$\cdot$ m$^2$}\\
    \hline
    \textbf{Value} & 0.028 & 0.046 & 6.4893 & 16.4562 & 29.5435\\
    \thickhline
    \end{tabular*}%
    }
\end{table}
 With some calculations, it can be verified that $\bm{P} = \bm{I}_{12}$ satisfies (\ref{eq:blockpineq}). For numerical simulation, we take the leader's trajectory tracking law, $\bm{u}_{\text{track}}$ (see Assumption \ref{asm:leadcont}), to be a trajectory-error minimizing control law, after the manner of \cite{enweremDistributedOptimalFormation2023a}. We also select a loop (figure eight) trajectory and set $\bm{x}_1(0) = \bm{x}_{1\text{rnd}}$, $\bm{x}_2(0) = \bm{x}_{2\text{rnd}}$, $\bm{x}_3(0) = \bm{x}_{3\text{rnd}}$, $\delta_{L1} = 1$ m, and $\delta_{12} = \delta_{13} = 0.8$ m. $\bm{x}_{1\text{rnd}}$, $\bm{x}_{2\text{rnd}}$, and $\bm{x}_{3\text{rnd}}$ are random initial state vectors for the first, second, and third followers, respectively.

\section{Numerical Results \& Discussion}
\label{sec:res}
\begin{figure*}[t]
     \centering
     \begin{subfigure}[t]{0.491\textwidth}
        \centering
        \includegraphics[trim=0 0 0 0,clip, width=0.8\columnwidth]{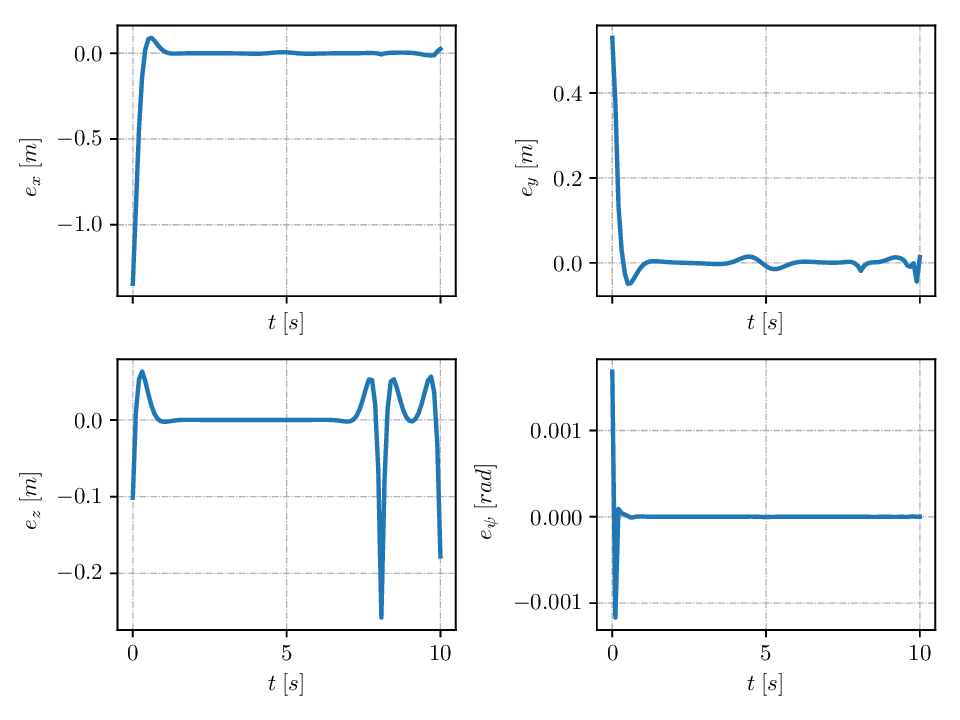}
        \caption{\label{fig:leaderrtrack}}
     \end{subfigure}
     \hfill
     \begin{subfigure}[t]{0.491\textwidth}
        \centering
        \includegraphics[trim=0 0 0 0,clip, width=0.8\columnwidth]{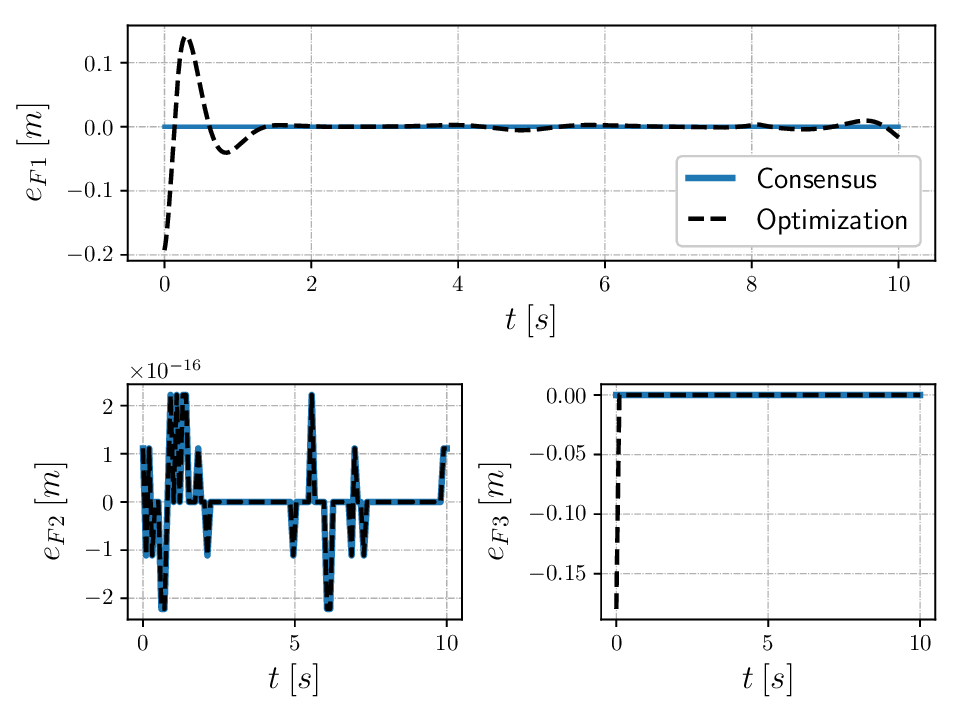}
        \caption{\label{fig:formerr}}
     \end{subfigure}
     \caption{(\subref{fig:leaderrtrack}) Time-evolution of the trajectory tracking error, given here by the errors (RMSE) in each state variable of the trajectory vector:  $e_x, e_y, e_z,$ and $e_\psi$. (\subref{fig:formerr}) Time-evolution of the formation error given by the RMSE between the desired and actual inter-agent distances, for the first ($e_{F1}$), second ($e_{F2}$), and third ($e_{F3}$) follower agents. Results obtained with consensus and those obtained via an exclusively optimization-based method \protect\cite{enweremDistributedOptimalFormation2023a} are also pictured.} %
\end{figure*}

As a prelude to discussing the consensus-based formation tracking results, Fig. \ref{fig:leadertraj} shows the trajectory tracking performance of the leader. Here, we see that the independent optimal control algorithm drives the leader (from an arbitrary initial state) so that it accurately tracks the desired trajectory, thus satisfying Assumption \ref{asm:leadcont}. In Fig. \ref{fig:consall}, the trajectory tracking and formation maintenance results are presented. Here, it is clear that under the consensus-derived control law, the trajectories of the followers closely track that of the leader resulting in tight trajectory tracking with formation persistence. Finally, to give a numerical sense of the formation tracking performance of our proposed control scheme, we present the tracking and formation errors of the MAS on Table \ref{tab:rmse}, with corresponding plots depicted on Figs. \ref{fig:leaderrtrack} and \ref{fig:formerr}. These errors are given in terms of the root-mean-square error (RMSE) between the leader and reference trajectories and the RMSE between the desired and actual inter-agent distances for each follower agent, respectively. As expected, the consensus protocol yields almost negligible formation errors, when viewed alone and also in comparison with errors obtained via a formation tracking approach based solely on optimization.

\begin{table}[htb]
    \caption{Tracking and Formation Errors (RMSE)}
    \label{tab:rmse}
    \centering
    \resizebox{\columnwidth}{!}{%
    \begin{tabular}{c c c c c}
    \thickhline
    \bf Agent & \bf Leader & \bf Follower 1 & \bf Follower 2 & \bf Follower 3\\
    \hline
    \bf RMSE (\textcolor{black}{consensus}) & 0.186 &  9.289$\times 10^{-17}$ & 7.692$\times 10^{-17}$ & 5.207$\times 10^{-17}$\\
    \bf RMSE (\textcolor{black}{optimization}) &  0.186 & 0.0302 & 7.692$\times 10^{-17}$ & 0.018\\
     \thickhline
    \end{tabular}%
    }
\end{table}

\section{Conclusions}
\label{sec:conc}
In sum, we presented results on consensus-based formation tracking for a general class of nonlinear systems --- control-affine systems with a state-dependent drift term --- and showed that, even with the MAS sharing information via a network topology encoded by a tree, precise formation tracking was still achieved. While our method delivers excellent formation tracking for a nonlinear system with a high-dimensional state space, such near-perfect performance is expected since we have assumed perfect knowledge of the agents' states. Thus, for the more interesting case where agents have uncertain dynamics, the development of similar control laws remains an open challenge. We look forward to exploring this direction in future research.

\section*{Acknowledgement}
C. Enwerem thanks Erfaun Noorani of the Electrical and Computer Engineering Department at the University of Maryland, College Park, for helpful technical discussions and comments on the content and presentation of this work.

\bibliographystyle{ieeetr}
\bibliography{ref}

\end{document}